\documentclass[aps,11pt,tightenlines,nofootinbib]{revtex4}
\usepackage{amsmath,amscd,amssymb}
\usepackage{color,epsfig}

\def\beq{\begin{equation}}
\def\eeq{\end{equation}}
\def\be{\begin{eqnarray}}
\def\ee{\end{eqnarray}}

\newcommand{\fract}[2]{\mbox{\small $\frac{#1}{#2}$}}

\begin{document}

\title{Axial current matrix elements 
and pentaquark decay widths in chiral soliton models}

\author{H. Weigel}

\affiliation{Fachbereich Physik, Siegen University,  
D--57068 Siegen, Germany}

\begin{abstract}
Here I explain why in chiral soliton models the hadronic transition operator 
of the pentaquark decay cannot be identified from the axial current.
\end{abstract}

\maketitle

\section{Introduction}
Many computations of pentaquark widths in soliton models fully rely on 
adopting the axial current as the transition operator for the hadronic 
decay $\Theta^+\to KN$~\cite{Diakonov:1997mm,Ellis:2004uz,addrefs}. 
These computations embody the $SU(3)$ generalization of the Goldberger--Treimann 
relation (GTR)\footnote{In the context of the Skyrme soliton model the GTR was first 
formulated in ref.~\cite{Adkins:1983ya}.} between the nucleon axial charge~($g_A$) 
and the pion nucleon coupling constant~($g_{\pi NN}$) to map the soliton model onto 
a Yukawa interaction. These calculations have been criticized for inconsistencies 
with the large $N_C$ limit~\cite{Itzhaki:2003nr}. In that limit the $\Theta$ has a 
non--zero mass gap to the nucleon and hence a non--zero width. On the other hand, 
the Skyrme model $KN$ phase shifts are exactly known for 
$N_C\to\infty$~\cite{Karliner:1986wq,Scoccola:1990pt}. They do not exhibit 
pronounced (narrow) resonances. More recently a detailed 
analysis~\cite{Walliser:2005pi} showed that these phase shifts indeed contain 
the pentaquark exchange contribution. Most crucially the transition matrix element 
for $\Theta^+\to NK$ was extracted and established that it does not equal the axial 
current matrix element suggested by the generlized GTR. Thus any chiral soliton 
model calculation of the $\Theta^+$ width that is based on identifying the 
transition matrix element from the axial current must be strongly doubted. 
Given that this identification is continuously employed in soliton motivated 
studies~\cite{Yang:2007yj,Diakonov:2006kh,Lorce:2006nq,Praszalowicz:2006rt,YKIS06} 
of pentaquark widths and that the resultant claim for the existence of narrow exotic 
baryons is airily adopted to this day~\cite{Kuznetsov:2007dy}, it occurs highly 
necessary to be emphatic on the arguments of ref.~\cite{Walliser:2005pi}.

Throughout I will discard flavor symmetry breaking. Though it is important for 
actual predictions to be reliable, it hides the main issue. Also, I will focus 
on the Skyrme model. Admittedly this model is insufficient in various aspects. 
Here the crucial point is the treatment of collective soliton excitations.
This is completely independent of the specific underlying effective meson 
theory. It is thus advantageous to consider the simplest model available.

\section{Decay widths from axial current matrix elements}

Models with explicit baryon ($B$) and meson ($\Phi$) fields commonly have 
tri--linear Yukawa interactions (the fields are multi--valued in flavor space), 
\beq
\mathcal{L}_{\rm int}=\frac{g_{\phi BB^\prime}}{M_B+M_B^\prime}\, \bar{\Psi}_{B} 
\gamma_5\gamma_\mu\left(\partial^\mu\Phi\right)\Psi_{B^\prime}\,.
\label{yukawa}
\eeq
The derivative interaction reflects chiral symmetry and $\gamma_5$ 
the  pseudoscalar nature of the considered meson. The Yukawa coupling 
leads to the standard width 
\beq
\Gamma(B^\prime\to B\Phi)=\frac{\overline{|\mathcal{M}|^2}}
{8\pi M_BM_{B^\prime}}\,|\vec{p}_\Phi|
\label{gamma}
\eeq
where $\mathcal{M}$ is the matrix element resulting from eq.~(\ref{yukawa}). 
The overbar denotes summing and averaging over spins. The 
details of this matrix element depend on the spins of the considered baryons. 
It suffices to keep in mind that $\mathcal{M}$ is linear in both, the coupling 
$g_{\phi BB^\prime}$ and the momentum of the final meson, $\vec{p}_\Phi$. 
The latter property results from the pseudoscalar nature of $\Phi$. So we 
have $\Gamma\propto g_{\phi BB^\prime}^2|\vec{p}_\Phi|^3$. The Yukawa model
reflects the GTR\footnote{Strictly speaking this relation is valid only at
zero momentum transfer and smoothness is assumed to extrapolate to the physical
point.} between the Yukawa coupling $g_{\pi NN}$ and the axial charge of the
nucleon, $g_A$
\beq
f_\pi g_{\pi NN} =M_N g_A\,.
\label{GTrel}
\eeq
This relation heavily relies on PCAC which expresses the non--conservation of
the axial current
\beq
\partial^\mu A_\mu^a(x)=f_a m_a^2 \phi^a(x)\,,
\label{PCAC}
\eeq
where $a$ is the flavor index.

In soliton models the situation is considerably different. Only meson fields 
are fundamental while baryons emerge as (topological) configurations thereof 
that solve the (classical) field equations. To study meson baryon interactions,
asymptotic meson states are constructed from small amplitude fluctuations about 
the soliton that describes the baryon. An immediate puzzle arises. Since the 
soliton is a stationary point, no term linear in the meson fluctuations exists. 
Hence there is no obvious coupling constant $g_{\phi BB^\prime}$ and profound 
assumptions are necessary to make use of eq.~(\ref{gamma}). The profound 
assumption often made in soliton models is to evaluate $g_A^{BB^\prime}$ (the 
axial current transition matrix element), use eq.~(\ref{GTrel}) to identify 
$g_{\phi BB^\prime}$ and substitute it into eq.~(\ref{gamma}) to compute the 
decay width. This is an attempt to map the soliton model onto the Yukawa model. 
Certainly, one must ask for the role of the GTR in soliton models. Before doing 
so, we will outline the computations of $g_A$, $g_{\pi NN}$ and its $SU(3)$ 
relatives from GTR.

Starting point is the hedgehog configuration
$U_0(\vec{x})={\rm exp}\left[i\hat{\vec{x}}\cdot\vec{\tau}F(r)\right]$,
that solves the classical field equations. In the next step collective 
coordinates $A(t)\in SU(3)$ are introduced via
\beq
U(\vec{x},t)=A(t)U_0(\vec{x})A^\dagger(t) \,.
\label{urot}
\eeq
Note that this configuration does not solve the stationary conditions,
eventually this gives rise to terms linear in the meson fields.
The $A(t)$ are treated quantum mechanically to generate states with good spin and 
flavor quantum numbers. Baryon wave--functions $\Psi_B(A)=\langle A|B\rangle$ 
emerge in the space of the collective coordinates. In the absence of flavor 
symmetry breaking these wave--functions are classified with respect
to $SU(3)$ flavor multiplets; spin $\fract{1}{2}$ states in the octet, 
anti--decuplet; spin $\fract{3}{2}$ in the decuplet; etc.\@.  This treatment 
is called the rigid rotator approach.

In the rigid rotator approach the axial current operator has the model 
independent from
\beq
A_i^a=\sum_{k=1,2,3}A^{(0)}_{ik}\,(\vec{x})D_{ak}+
\sum_{\genfrac{}{}{0pt}{}{k=1,2,3}{\alpha,\beta=4,\ldots,7}}
A^{(1)}_{ik}(\vec{x})\,d_{k\alpha\beta}D_{a\alpha}R_\beta+
\sum_{k=1,2,3}A^{(2)}_{ik}(\vec{x})\,D_{a8}R_k
\label{axialcurrent}
\eeq
up to omitted flavor symmetry breaking. The structure of the coefficient 
functions is 
$A^{(m)}_{ik}(\vec{x})=A_1^{(m)}(r)\delta_{ik}+A_2^{(m)}(r)\hat x_i\hat x_k$.
The $A_{1,2}^{(m)}(r)$ are radial functions through the profile function 
$F(r)$. The $D_{ab}={\scriptstyle \frac{1}{2}}{\rm tr}
\left(\lambda_a A \lambda_b A^\dagger\right)$ 
and the $R_a$ are the adjoint representation
of the $SU(3)$ collective coordinates and the intrinsic $SU(3)$ generators,
respectively. It is legitimate to use isospin invariance and compute
$g_A$ as the nucleon matrix element $\langle 2A_3^3\rangle$. Then 
eq.~(\ref{GTrel}) implies~\cite{Diakonov:1997mm,Ellis:2004uz}
\beq
g_{\pi NN}=\frac{7}{10}\left[G_0+\frac{1}{2}G_1+\frac{1}{14}G_2\right]
\qquad{\rm with}\qquad
G_m=-\frac{8\pi M_N}{3f_\pi}\int_0^\infty dr r^2\left[
A_1^{(m)}(r)+\frac{1}{3}A_2^{(m)}(r)\right]\,.
\label{gpinn}
\eeq
The relative coefficients stem from the nucleon matrix elements of the 
collective coordinate operators in eq.~(\ref{axialcurrent}). They are 
readily obtained from $SU(3)$ Clebsch--Gordan coefficients, {\it e.g.\@}
$\langle p\uparrow |D_{33}|p\uparrow\rangle=-7/30$. Generalizing the above 
result for $g_{\pi NN}$ to flavor $SU(3)$ yields coupling constants 
\beq
G_{10}=G_0+\frac{1}{2}G_1
\qquad {\rm and} \qquad
G_{\overline{10}}=G_0-G_1-\frac{1}{2}G_2
\label{G1010bar}
\eeq
that (under the GTR assumption) respectively measure the coupling of baryons from 
the decuplet ($\Delta$) and the anti--decuplet ($\Theta^+$) to those in the octet 
(nucleon, hyperons). These coupling constants enter the matrix element $\mathcal{M}$ 
and predict widths for hadronic baryon decays via eq.~(\ref{gamma}):
$\Gamma(\Delta\to N\pi)\propto G_{10}^2|\vec{p}_\pi|^3$ and
$\Gamma(\Theta^+\to NK)\propto G_{\overline{10}}^2|\vec{p}_K|^3$. The omitted
constants of proportionality are merely kinematical factors~\cite{misconduct}. 
Model calculations~\cite{Kanazawa:1987uv,Blotz:1994wi,Weigel:1995cz} indicate that 
$G_0$ and $G_1$ are comparable. That is, significant cancellations cause
$G_{\overline{10}}$ to be rather small. This has been the main argument for 
claiming a $\Theta^+$ width of the order of only a few ${\rm MeV}$, or even less. 
The cancellations between $G_0$ and $G_1$ persist when the number ($N_C$) of 
color degrees of freedom is sent to infinity~\cite{Praszalowicz:2003tc}. This 
completes the way of thinking about pentaquark decay widths put forward in 
refs.~\cite{Diakonov:1997mm,Ellis:2004uz} and frequently adopted later 
on~\cite{addrefs}. A couple of issues doubt this approach already afore we 
test it against incontrovertible results from the phase shift analysis:
\begin{itemize}
\item
The classical field equations affect only the first part
$\partial_i A^{(0)}_{ik}=\mathcal{O}\left(m_\pi^2\right)$ while the last term 
($A^{(2)}$) vanishes or is at least small because it essentially is the axial 
singlet matrix element. On the other hand $\partial_i A^{(1)}_{ik}$ is not part 
of any equation of motion. Hence the axial current computed solely 
from the classical profile functions violates PCAC~\cite{Weigelappendix}. As a 
consequence, the use of GTR in $SU(3)$ soliton models is questionable because
a major entry is not met.
\item
The above derivation only involves the classical soliton and there is no 
reference to asymptotic meson states. In two flavor soliton models the GTR
arises from the long range behavior of the soliton profile~\cite{Adkins:1983ya} 
and has been identified from one--pion exchange contribution to the nucleon--nucleon 
interaction. However, this process does not require asymptotic pion states. Also, 
that argument strongly relies on pions being massless. For $m_\pi>0$, $g_A$ cannot 
be read off from the long range behavior and thus not be related to $g_{\pi NN}$.
\end{itemize}
It is thus not surprising that $SU(3)$ Skyrme model calculations severely 
fail to reproduce GTR when $g_{\pi NN}$ is identified from the long range 
behavior of the soliton~\cite{Kanazawa:1987uv}. Evidently, it is
not possible to directly map soliton models onto the Yukawa model.

\section{Rotation--vibration coupling and $KN$ scattering}

\begin{figure}[t]
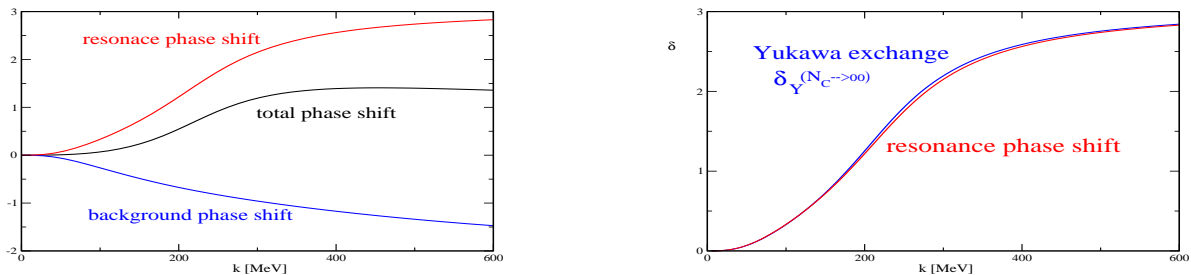

\centerline{
\epsfig{file=resonance.eps,height=4.5cm,width=7.0cm}\hspace{2cm}
\epsfig{file=yukawa.eps,height=4.5cm,width=7.0cm}}
\caption{\label{fig:phase}Skyrme model results for momentum dependent
phase shifts in the $\Theta^+$ channel. The 'total phase shift' is the 
result in adiabatic approximation, the 'background phase shift' 
describes scattering in the space orthogonal to the soliton's rigid rotation
and the 'resonance phase shift' is their difference, compared to the Yukawa 
exchange contribution~(\ref{resphase},\ref{widthfunction}) in the 
right panel. The pictures are adopted from ref.~\cite{Weigel:2006vt}.}
\end{figure}

In principle, it must be possible to extract $\Theta^+$ properties from kaon 
nucleon scattering data. After all, that is the process in which resonances 
are to be observed. This process can be studied within a given soliton
model \emph{without} reference to the Yukawa model. The corresponding phase shifts 
have been computed in the Skyrme model~\cite{Karliner:1986wq,Scoccola:1990pt}
within the so--called adiabatic approximation, which neglects the dynamical properties 
of the collective modes. This model treatment is exact to first non--trivial order 
in the large--$N_C$ expansion. A typical Skyrme model result is shown as 
\emph{total phase shift} in figure~\ref{fig:phase}.
Since this is exact to $\mathcal{O}\left(N_C^0\right)$, any treatment (that might 
include subleading pieces) of collective degrees of freedom in the Skyrme model 
\emph{must} reproduce this phase shift in the limit $N_C\to0$. This concerns the 
full momentum dependence and not only an attempt to match a single 
parameter~\cite{Diakonov:2006kh}.

From figure~\ref{fig:phase} the immediate question arises whether or 
not the pentaquark channel resonates as $N_C\to\infty$. If at all, this
concerns the collective modes, eq.~(\ref{urot}). As a first response to
this question, one may constrain the small amplitude fluctuation to be
orthogonal to the collective modes. The phase shift computed from these
restricted fluctuations is shown as the \emph{background phase shift}
in figure~\ref{fig:phase}. Its difference to the total phase shift 
defines the \emph{resonance phase shift}. Obviously the latter resonates,
though it is definitely not narrow. Of course, the challenge is to verify 
that this resonance phase shift arises from the exchange of the collective 
excitation $\Theta^+$. Then collective modes must be treated dynamically 
within the scattering problem. This was done in ref.~\cite{Walliser:2005pi} 
by considering vibrations ($\widetilde{\eta}$) about the rotating hedgehog, 
(The interested reader should consult that paper for quantitative results,
particularly for the realistic case $N_C=3$ and $m_K\ne m_\pi$.). 
The configuration~(\ref{urot}) does not solve the stationary conditions, this 
will now give rise to terms linear in $\widetilde{\eta}$ that couple to the
collective coordinates and their time derivative. The constraints that 
ensure $\widetilde{\eta}$ to be orthogonal to the collective modes, yield
additional linear terms. After quantization, the linear terms contain only 
a \emph{single} collective coordinate operator 
\beq
\hat{X}_{ak}=\sum_{\alpha,\beta=4,\ldots,7}d_{k\alpha\beta}D_{a\alpha}R_\beta\,.
\label{singleoperator}
\eeq
This operator also occurs in the axial current operator,
eq.~(\ref{axialcurrent}). In the limit $N_C\to\infty$ the Schr\"odinger equation 
for $\widetilde{\eta}$ has a very simple solution~\cite{Walliser:2005pi}:
$|\widetilde{\eta}\rangle=|\eta\rangle- |z\rangle \langle \eta |z\rangle$,
where $\langle \vec{x}|\eta\rangle$ is the wave function
in the adiabatic approximation and $\langle \vec{x}|z\rangle$
is the (properly normalized) wave function that represents the collective
modes. Since $\langle \vec{x}|z\rangle$ is determined by the soliton 
configuration it is localized in space. Thus $\langle \vec{x}|\eta\rangle$ 
and $\langle \vec{x}|\widetilde{\eta}\rangle$ behave identically in the 
asymptotic regime and $\widetilde{\eta}$ indeed reproduces the 
\emph{total phase shift} of figure~\ref{fig:phase} as $N_C\to\infty$. To 
extract the information about the collective excitations that is contained
in $\widetilde{\eta}$, it is fruitful to introduce fluctuations $\overline{\eta}$ 
which are the solution to the Schr\"odinger equation with $\hat{X}\to0$,
{\it i.e.\@} \underline{without} coupling to the collective coordinates. The
so--generated equation of motion is that from the adiabatic approximation 
augmented by the constraints. Hence the $\overline{\eta}$ phase shift is the 
background phase shift in figure~\ref{fig:phase}. In the full $\widetilde{\eta}$
problem the effect of non--zero $\hat{X}$ is to add the resonance phase shift 
$\delta_Y(k)$ to the $\overline{\eta}$ phase shift with
\beq
{\rm tan}\left(\delta_Y(k)\right)=\frac{\Gamma(\omega_k)/2}
{\omega_\Theta-\omega_k+\Delta(\omega_k)}\,.
\label{resphase}
\eeq
Here $\omega_\Theta$ is the excitation energy of the pentaquark as computed in 
the rigid rotator approach while $\Delta$ denotes the energy shift. Numerically 
$\Delta$ turns out to be negligibly small. This shows that the rigid rotator 
approach reliably predicts the pentaquark mass (in a model)~\cite{sigma}. 
The width function
\beq
\Gamma(\omega_k)=2k\omega_0X_\Theta^2
\left|\langle\overline{\eta}_{\omega_k}|(2\lambda)|z\rangle\right|^2
\label{widthfunction}
\eeq
describes the (Yukawa) exchange contribution of a pentaquark to kaon--nucleon 
scattering. Here, the explicit expressions for the normalization factor $\omega_0$ 
and the radial function $\lambda$ are of minor importance. For $N_C\ne3$ the
low--lying $SU(3)$ representations are no longer octet, decuplet, anti--decuplet 
etc.\@. This induces $N_C$ dependences for $\Psi_A$, the energy eigenvalues such 
as $\omega_\Theta$, and the matrix element $X_\Theta=\sqrt{\frac{32}{N_C}}
\langle\Theta^+\uparrow|X_{43}+iX_{53}|n\uparrow\rangle$ turns into a function of 
$N_C$. It is normalized such that $\lim_{N_C\to\infty}X_\Theta=1$. 

\section{Comparison and Critique}

The right panel of figure~\ref{fig:phase} shows the Yukawa--exchange phase shift
as numerically computed from eq.~(\ref{widthfunction}) for $N_C\to\infty$. Obviously 
and most importantly it exactly matches the resonance phase shift! Unambiguously
$\Gamma(\omega_k)$ is the correct width function in this model (at least for 
$N_C\to\infty$). It is evidently very different from the width function computed via 
GTR from the axial current. Most remarkably $\Gamma(\omega_k)$ contains only a 
\emph{single} collective coordinate operator. Thus there cannot be any 
cancellation that would yield a small width. 

There is a self--explanatory and rigorous reason for the appearance of only a single 
collective coordinate structure in the transition operator. To make contact with 
the adiabatic approximation (that is exact as $N_C\to\infty$), the equations of 
motion for the fluctuations are solved in the body--fixed frame, wherein
the fluctuations rotate along with the soliton, eq.~(\ref{urot}). In these 
equations the collective coordinates can only show up via the angular velocities, 
$\Omega_a=-i{\rm tr}\left[\lambda_aA(t)\dot{A}^\dagger(t)\right]$. Upon 
quantization the $\Omega_a$ are replaced by the generators $R_a$. Without
the $D_{ab}$ available, there is only one possible kaon $P$--wave coupling
which is Hermitian and behaves properly under $SU(3)$: 
$\sum \hat{x}_k\,d_{k\alpha\beta}\eta_\alpha R_\beta$. Subsequently matrix elements 
for the lab--frame fluctuations $\xi_a=D_{ab}\eta_b$ are required. This leads 
to $\hat{X}$ as the only allowed operator. Since this argument is irrespective
of the considered chiral soliton model, the emergence of only a single 
collective coordinate operator for the hadronic transition $\Theta\to KN$ is 
common to all chiral soliton models.

The detailed analysis~\cite{Walliser:2005pi} 
of $\Gamma(\omega_k)$ reveals a few more discrepancies to the axial current 
approach. The $|\vec{p}_K|^3$ behavior is seen only in the energy regime 
slightly above threshold; at larger energies is levels off. Though 
the correct width function does definitely not contain a $G_0$ type piece, 
it it seems plausible to identify the $G_1$ contribution
with $\Gamma(\omega_k)$ in the plane wave approximation in the $\vec{p}_K\to0$
limit because it contains the same collective coordinate operator. However, 
the actual computation shows that the integrands of the spatial integrals
differ by a factor ${\rm cos}(F/2)$. 

When symmetry breaking is included, the $\Lambda$ channel must be 
incorporated to reproduce the correct total phase shift when $N_C\to\infty$. 
Also an additional collective coordinate operator 
$\hat{Y}_{ak}=\sum_{\alpha,\beta=4,\ldots,7}d_{k\alpha\beta}D_{a\alpha}D_{8\beta}$ 
emerges. In the large--$N_C$ limit it behaves similarly to the $G_0$ type piece, 
but in general no relation can be made.  

\section{A note on $\Delta$}

Many approaches describe the width of the $\Delta$ resonance via Yukawa 
interaction in pion nucleon scattering; thereby generalizing the GTR. 
This stimulates to discuss the consequences of the above results 
for the soliton model description of the $\Delta$ width. 

The treatment that consistently describes the $\Theta$
width is characterized by essentially two features. First, the space
of fluctuations is parted into a piece that contains collective modes 
and its orthogonal subspace. Only the latter contains to scattering meson
states. For pion nucleon scattering this partition is not as problematic 
as for three flavor processes because the Wess--Zumino term can be ignored. 
Then terms that are subleading in the $1/N_C$ expansion and quadratic in 
the fluctuations emerge that couple these two subspaces. They are thus linear 
in the small amplitude fluctuations that describe asymptotic pion fields.
Second, the collective modes are integrated out similar to the Lee--model
approach~\cite{Henley:1962}. This induces a separable potential for
the fluctuations in the orthogonal subspace. Treated in an $R$--matrix
formalism, this potential yields the resonance phase shift. As is deduced
from the pentaquark problem the $R$--matrix elements must be evaluated from 
fluctuation wave--functions that are distorted by the classical soliton; 
the plain wave approximation is inconsistent with the partitioning of the 
fluctuation space. So far this agenda has not been fully carried out for the 
$\Delta$ resonance. The study of chapter~10 in ref.~\cite{Schwesinger:1988af} 
seems to come closest. Even though those results for the pion nucleon 
scattering amplitude in the $\Delta$ channel agree reasonably well with 
data, it should be stressed that they are obtained in the plain wave 
approximation, just criticized. Interestingly enough this approach does 
\emph{not} generate a $\pi NN$ vertex~\cite{Saito:1987uh,Holzwarth:1990ym},
irregardless of the plain wave approximation. Stated otherwise, the scenario 
that potentially describes the $\Delta$ resonance well, or at least consistently 
in a given soliton model, does not alter the pion nucleon coupling constant 
that is basic to GTR, eq.~(\ref{GTrel}). So, as in the pentaquark channel, the 
soliton model computation of the $\Delta$ width does not proceed by generalizing
the GTR\footnote{The study of ref.~\cite{Verschelde:1989sn} finds agreement
between the Skyrme and isobar model $T$--matrices in the $P$--wave channels at low
energies, Yet, this agreement cannot be attributed to individual nucleon or 
$\Delta$ Yukawa couplings.}.
This is no contradiction to fundamental concepts of hadron physics 
because sandwiching the PCAC relation~(\ref{PCAC}) between states other than
the nucleon (and its octet partners) is impossible without assumptions about 
the nature of these states. For example, regarding $g_{\pi NN}$ and $g_{\pi N\Delta}$ 
with equal rigor, implies that the $\Delta$ would be an asymptotic 
state, yet it is a resonance. We note, however, that the GTR is indeed
reproduced in the soliton model description of the nucleon--nucleon
potential~\cite{Adkins:1983ya,Jackson:1985bn}.

\section{Conclusions}

Here I have argued that pentaquark widths may not be computed from axial 
current matrix elements in chiral soliton models. The model prediction for 
the kaon nucleon phase shift is known in limit $N_C\to\infty$ and the axial 
current approach evidently fails to reproduce it. Though this fact is known for 
some time, this short discussion occurred necessary because this erroneous 
identification keeps on being applied. In chiral soliton models 
pentaquark widths (probably neither those of other baryon resonances) should 
not be estimated by mapping onto the Yukawa model via the GTR. Since the 
so--computed pentaquarks widths are not reliable predictions, the non--observation 
of such a narrow resonance (which seems more or less certain by 
now~\cite{Hicks:2007vg}) should not be used against the chiral soliton picture for 
baryons. The statement that chiral soliton models predict a \emph{very narrow}
pentaquark baryon in the $S=+1$ channel essentially is a myth. 

\end{document}